\documentclass[a4paper,12pt]{article}
\usepackage{graphicx}
\usepackage{amssymb}
\usepackage{a4}

\title{The quantum Neumann model: refined semiclassical results.}
\date{September 2005}
\author{Marc Bellon and Michel Talon\thanks{LPTHE, CNRS et Universit\'es
Paris VI--Paris VII (UMR 7589),
 Bo\^{\i}te 126,
 4 place Jussieu, F-75252 PARIS CEDEX 05}
}

\begin{document}

\maketitle

\begin{abstract}
We extend the semiclassical study of the Neumann model down to
the deep quantum regime. A detailed study of connection formulae
at the turning points allows to get good matching with the exact
results for the whole range of parameters.
\end{abstract}
\vfill
LPTHE--05--29
\eject

\section{Introduction}

The Neumann model is an interesting example of integrable model, both
classically and quantum mechanically, which describes the motion of a point on
a sphere subject to a harmonic potential. In the quantum case a lot
of effort has been devoted to study the semi--classical
approximation~\cite{Gurarie}, since it deals with algebro--geometric objects
which appear naturally in the classical theory. However this study becomes
more interesting and concrete when one has precise numbers to state 
effectively the Bohr-Sommerfeld conditions and compare them with exact
spectra. In a previous works~\cite{BT05,BT05a},
we showed how to solve numerically the separated Schr\"odinger equation to
obtain numerical values of the energy and compared this to a semiclassical
computation valid for a large radius of the sphere. This case corresponds to
the localization of the particle around the antipodal minima of the potential.
These localized states come in pairs which are split by the tunneling
probability between the two poles. The two states are either symmetric or
antisymmetric under exchange of the poles. 

When the sphere radius shrinks, the states become delocalized and the
degeneracy is lifted by a finite quantity. In the limit of zero radius, the
potential becomes irrelevant and the energy is the one of the free particle on
the sphere $1/2j(j+1)$. The symmetric and antisymmetric solutions go to states
with $j$ differing by one. When flowing between the large radius and small
radius cases, the matching conditions used in the WKB analysis must change, in
particular to explain this degeneracy lifting. We study in this work the
evolution of the matching phase which either goes from $\pi/4$ to 0 through
$\pi/8$ for a symmetric wave function of from $\pi/4$ to $\pi/2$ through
$3\pi/8$ for an antisymmetric wave function.

In a first part, we study the zero radius limiting case, where the energy is
reproduced by the WKB method up to a small constant. Then we detail the
numerical evolution of the phase. In a last part, we explain this evolution
through the asymptotic analysis of a simplified model.

\section{Semiclassical analysis of spheroidal harmonics.}

Even if the Neumann model has been generalized to arbitrary dimensions, there
are no really new phenomena appearing above the three dimensions to which we
limit ourselves in this work. The parameters of the Neumann model are the
oscillator strengths which can be reduced without loss of generality to be 0, 1
and $y$. The Schr\"odinger equation separates using the Neumann coordinates,
yielding the one-dimensional equation:
\begin{equation}
\biggl( {d^2\over dt^2} +{1\over 2}\Bigl({1\over t} + {1\over t-1}+ {1\over t-y}
\Bigr)  {d\over dt} -\Bigl({v-f_2-f_3\over t} + {f_2\over t-1}+ {f_3\over t-y}
\Bigr) \biggr) \Psi(t) = 0 \label{wangerin}
\end{equation} 
In this equation, $f_2$ and $f_3$ are the eigenvalues of the conserved
quantities which have to be determined and $v$ is $r^2/(4\hbar^2)$ where $r$ is
the radius of the sphere. The energy is $E= 2\hbar^2 (f_2+y f_3)$. In the
sequel, we take $\hbar=1$. The points 0, 1 and $y$ are regular singularities
of eq.~(\ref{wangerin}) with exponents $0$ and $1/2$, corresponding to
solutions with monodromies $\pm1$. In order to recover a well defined solution
on the sphere with definite parity properties under the three possible
reflexions, the solution must have definite monodromies at the three
singularities~\cite{BT05}. An even solution is thus a solution with monodromy 1
at 0, 1 and~$y$, that is a function which is analytic on the whole complex plane. If we
want an odd solution under $x_1\to -x_1$, we factor $\sqrt t$ in $\Psi$ and
search for an analytic solution to the equation:
\begin{eqnarray}
&&\hskip -1cm \biggl( {d^2\over dt^2} +{1\over 2}\Bigl({3\over t} + {1\over t-1}
+ {1\over t-y} \Bigr)  {d\over dt} -\nonumber\\
&&\Bigl({v-f_2-f_3+1/4+1/(4y)\over t} +
{f_2-1/4\over t-1} + {f_3-1/(4y)\over t-y} \Bigr) \biggr) \Psi(t) = 0
\label{wangerin2}
\end{eqnarray}
Similar equations can be written for the six other possible combinations of
parities.

In our previous work~\cite{BT05a}, we showed how the semiclassical analysis
nicely fits the exact numerical spectra of the Neumann model in the large $v$
limit, when the point is confined around the poles. To extend the analysis to
the whole range of values for $v$, we first consider the $v=0$ case,
corresponding to the spheroidal harmonics. The energy is known exactly in this
case, since it is simply the energy of a free particle on the sphere,
${1\over2}j(j+1)$.

In this case, the potential simplifies since the numerator is of degree one:
\begin{equation}
\label{PotSph} V= -{E \over 2} { t-b \over t(t-1)(t-y) }
\end{equation} 
The Bohr--Sommerfeld integrals are thus elliptic integrals, with the four
special points 0, 1, $b$ and $y$ corresponding to the vertices of the period
rectangle, and a pole at infinity. We have two cases to study according to the
position of $b$ with respect to 1.

The quantification conditions, in the case with all monodromies equal to 1, are
either ones of:
\begin{eqnarray}
b>1, \qquad \int_0^1 \sqrt V dt &=& m \pi, \qquad \int_b^y \sqrt V dt = (n+{1\over4}) \pi
\label{qan1}\\
b<1, \qquad \int_0^b \sqrt V dt &=& (m + {1\over 4}) \pi, \qquad \int_1^y \sqrt V dt = n \pi
\label{qan2}
\end{eqnarray} 
These quantification conditions involve the usual $\pi/4$ phase factor coming
from the asymptotics of the Airy function around the turning point $b$, see
e.g.~\cite{Pau}.

Adding the two Bohr--Sommerfeld integrals, one gets a condition which does not
depend on the position of $b$ with respect to 1. In fact, this sum can be
computed and does not depend on $b$. Indeed, the integrand is an abelian
integral of the third type with simple poles at the two points over $t=\infty$.
The real part of the integral on the real axis is just the sum we want to
evaluate, since the integrand is purely imaginary on the complementary
segments. Hence the residue theorem allows to calculate this sum. The residue
is $i\sqrt{E/2}$ and since the integration contour runs through the pole, we
get a contribution of $i \pi$ times the residue. The final quantification
condition is therefore:
\begin{equation}
\pi \sqrt{E/2} = ( m+n+{1\over 4} ) \pi
\end{equation} 
The semiclassical energy thus depends only on the total number of excitations
and not on the individual values of $m$ and $n$. The angular momentum is $j =
2(m+n)$ so that $E = 1/2 (j + 1/2)^2 = 1/2 j (j+1) + 1/8$. This differs by
merely $1/8$ from the exact result and is exactly the value of $L^2$ which must
be used in a semiclassical treatment of spherically symmetric potential
according to Langer~\cite{Lan38}. 

When $b$ varies, we should be able to pass smoothly from the case of
eq.~(\ref{qan1}) to the one of eq.~(\ref{qan2}). The additional phases
therefore go from $0$ to $\pi/4$ and from $\pi/4$ to 0, but their sum should
remain constant for the semiclassical energy to be completely independent of
$b$. In particular, in the case $b=1$, we should have for symmetry reason the
same $\pi/8$ phase on both sides. We shall show below that this is indeed the
case.

\section{Numerical investigations.}
%GNUPLOT: LaTeX picture with Postscript
\begin{figure}
\begin{picture}(0,0)%
\includegraphics{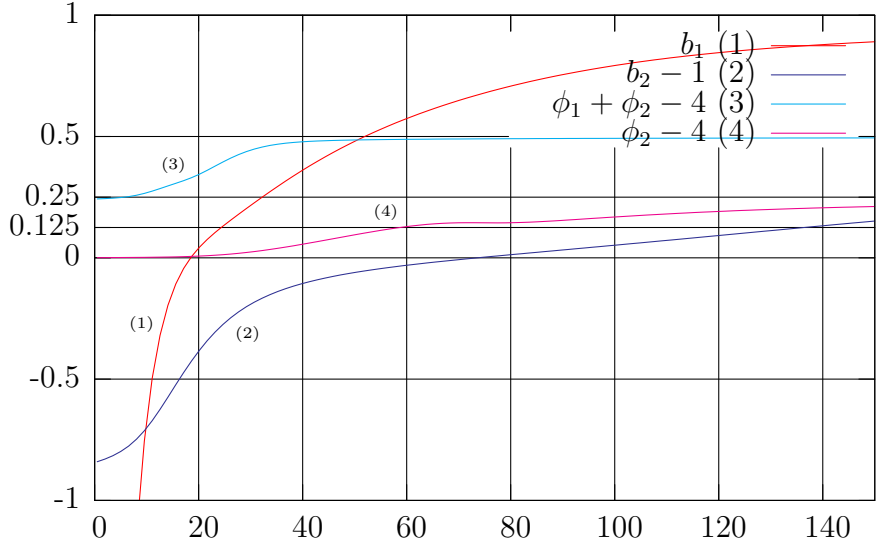}%
\end{picture}%
\begingroup
\setlength{\unitlength}{0.0200bp}%
\begin{picture}(18000,10800)(0,0)%
\put(2200,10250){\makebox(0,0)[r]{\strut{} 1}}%
\put(2200,7963){\makebox(0,0)[r]{\strut{} 0.5}}%
\put(2200,6819){\makebox(0,0)[r]{\strut{} 0.25}}%
\put(2200,6247){\makebox(0,0)[r]{\strut{} 0.125}}%
\put(2200,5675){\makebox(0,0)[r]{\strut{} 0}}%
\put(2200,3388){\makebox(0,0)[r]{\strut{}-0.5}}%
\put(2200,1100){\makebox(0,0)[r]{\strut{}-1}}%
\put(4200,7500){\makebox(0,0)[r]{\strut{}\tiny (3)}}%
\put(3600,4500){\makebox(0,0)[r]{\strut{}\tiny (1)}}%
\put(5600,4300){\makebox(0,0)[r]{\strut{}\tiny (2)}}%
\put(8200,6600){\makebox(0,0)[r]{\strut{}\tiny (4)}}%
\put(2475,550){\makebox(0,0){\strut{} 0}}%
\put(4435,550){\makebox(0,0){\strut{} 20}}%
\put(6395,550){\makebox(0,0){\strut{} 40}}%
\put(8355,550){\makebox(0,0){\strut{} 60}}%
\put(10315,550){\makebox(0,0){\strut{} 80}}%
\put(12275,550){\makebox(0,0){\strut{} 100}}%
\put(14235,550){\makebox(0,0){\strut{} 120}}%
\put(16195,550){\makebox(0,0){\strut{} 140}}%
\put(14950,9675){\makebox(0,0)[r]{\strut{}$b_1$ (1)}}%
\put(14950,9125){\makebox(0,0)[r]{\strut{}$b_2-1$ (2)}}%
\put(14950,8575){\makebox(0,0)[r]{\strut{}$\phi_1+\phi_2-4$ (3)}}%
\put(14950,8025){\makebox(0,0)[r]{\strut{}$\phi_2-4$ (4)}}%
\end{picture}%
\endgroup
\caption{Variation of $b_1$, $b_2$ and phases as function of $v$.\label{abs}}
\end{figure}

In order to study the variation of the phases appearing in the connection
formulae at the turning points, we numerically solve the
equation~(\ref{wangerin}) using the methods of~\cite{BT05}. From the values of
$f_2$ and $f_3$, we calculate the zeros $b_1$ and~$b_2$ of the potential, which
can be written:
\begin{equation}
	V = -v{(t-b_1)(t-b_2) \over t(t-1)(t-y)} 
\end{equation} 
%GNUPLOT: LaTeX picture with Postscript
\begin{figure}
\begin{picture}(0,0)%
\includegraphics{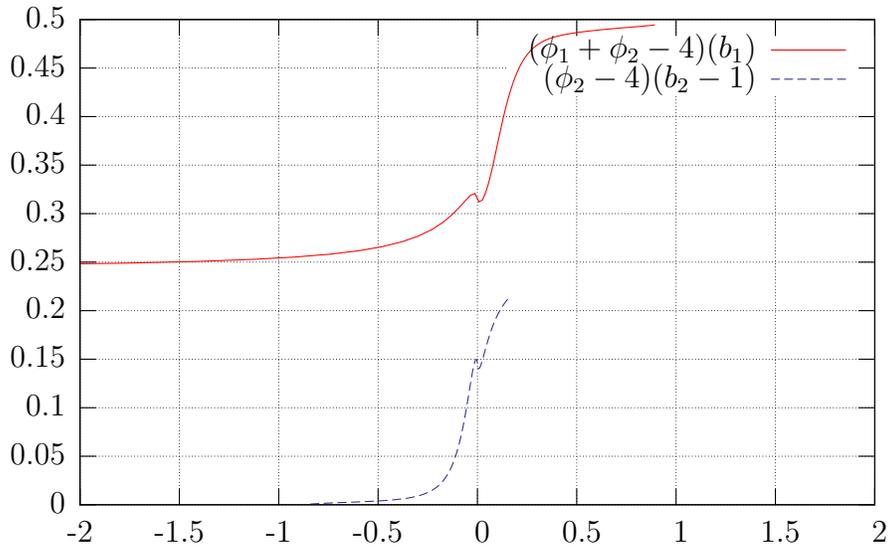}%
\end{picture}%
\begingroup
\setlength{\unitlength}{0.0200bp}%
\begin{picture}(18000,10800)(0,0)%
\put(1925,1100){\makebox(0,0)[r]{\strut{} 0}}%
\put(1925,2015){\makebox(0,0)[r]{\strut{} 0.05}}%
\put(1925,2930){\makebox(0,0)[r]{\strut{} 0.1}}%
\put(1925,3845){\makebox(0,0)[r]{\strut{} 0.15}}%
\put(1925,4760){\makebox(0,0)[r]{\strut{} 0.2}}%
\put(1925,5675){\makebox(0,0)[r]{\strut{} 0.25}}%
\put(1925,6590){\makebox(0,0)[r]{\strut{} 0.3}}%
\put(1925,7505){\makebox(0,0)[r]{\strut{} 0.35}}%
\put(1925,8420){\makebox(0,0)[r]{\strut{} 0.4}}%
\put(1925,9335){\makebox(0,0)[r]{\strut{} 0.45}}%
\put(1925,10250){\makebox(0,0)[r]{\strut{} 0.5}}%
\put(2200,550){\makebox(0,0){\strut{}-2}}%
\put(4072,550){\makebox(0,0){\strut{}-1.5}}%
\put(5944,550){\makebox(0,0){\strut{}-1}}%
\put(7816,550){\makebox(0,0){\strut{}-0.5}}%
\put(9688,550){\makebox(0,0){\strut{} 0}}%
\put(11559,550){\makebox(0,0){\strut{} 0.5}}%
\put(13431,550){\makebox(0,0){\strut{} 1}}%
\put(15303,550){\makebox(0,0){\strut{} 1.5}}%
\put(17175,550){\makebox(0,0){\strut{} 2}}%
\put(14950,9675){\makebox(0,0)[r]{\strut{}$(\phi_1+\phi_2-4)(b_1)$}}%
\put(14950,9125){\makebox(0,0)[r]{\strut{}$(\phi_2-4)(b_2-1)$}}%
\end{picture}%
\endgroup
\caption{Phases as a function of the roots $b_i$. \label{rel}}
\end{figure}%
The Bohr--Sommerfeld integrals are taken on intervals where the potential is
positive, with boundary points in the set $\{0,1,y,b_1,b_2\}$. We numerically
evaluate them and plot their differences with naive expectations in units of
$\pi$. In the following plots, we study the case where the excitation numbers
are $n=0$ in the interval $[0,1]$ and $m=4$ in the interval $[1,y=2]$. We
obtain similar results for other values of the parameters $n$, $m$ and $y$.
The plot (Figure \ref{abs}) clearly shows that the additional phase in
the interval $[1,2]$ goes from 0 when $b_2$ is smaller than 1 to a value
approaching $\pi/4$ in the other case, with the variation stronger when $b_2$
crosses 1. To isolate the dependence on $b_1$, we plot the sum of the two
integrals which should not depend on the position of $b_2$ with respect to 1. It
varies from $\pi/4$ for negative $b_1$ to $\pi/2$ for positive $b_1$.

In order to show the details of the transition region, we next 
plot (Figure \ref{rel}) the relevant phases as functions of $b_1$ or $b_2-1$. We
see that the dependence of the phase is not monotonic, but there is a small "glitch"
just at $b_1=0$ or $b_2=1$. We shall see below that this can indeed be
explained by analytical
studies. We also see that the value of the phase is not precisely $\pi/8$ at
this point of transition.

%GNUPLOT: LaTeX picture with Postscript
\begin{figure}
\begin{picture}(0,0)%
\includegraphics{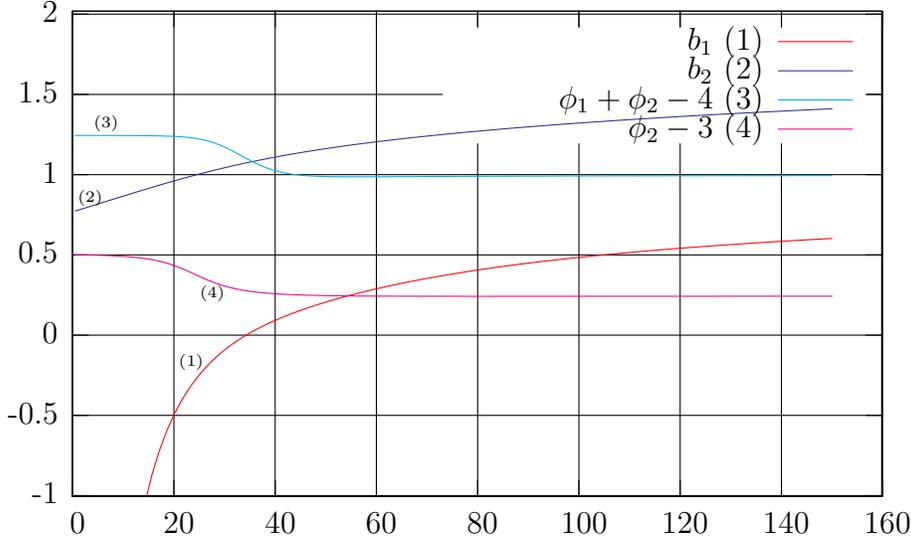}%
\end{picture}%
\begingroup
\setlength{\unitlength}{0.0200bp}%
\begin{picture}(18000,10800)(0,0)%
\put(1650,1114){\makebox(0,0)[r]{\strut{}-1}}%
\put(1650,2627){\makebox(0,0)[r]{\strut{}-0.5}}%
\put(1650,4140){\makebox(0,0)[r]{\strut{} 0}}%
\put(1650,5653){\makebox(0,0)[r]{\strut{} 0.5}}%
\put(1650,7166){\makebox(0,0)[r]{\strut{} 1}}%
\put(1650,8679){\makebox(0,0)[r]{\strut{} 1.5}}%
\put(1650,10193){\makebox(0,0)[r]{\strut{} 2}}%
\put(2800,8200){\makebox(0,0)[r]{\strut{}\tiny (3)}}%
\put(4400,3700){\makebox(0,0)[r]{\strut{}\tiny (1)}}%
\put(2500,6800){\makebox(0,0)[r]{\strut{}\tiny (2)}}%
\put(4800,5000){\makebox(0,0)[r]{\strut{}\tiny (4)}}%
\put(1929,550){\makebox(0,0){\strut{} 0}}%
\put(3837,550){\makebox(0,0){\strut{} 20}}%
\put(5745,550){\makebox(0,0){\strut{} 40}}%
\put(7652,550){\makebox(0,0){\strut{} 60}}%
\put(9560,550){\makebox(0,0){\strut{} 80}}%
\put(11468,550){\makebox(0,0){\strut{} 100}}%
\put(13376,550){\makebox(0,0){\strut{} 120}}%
\put(15284,550){\makebox(0,0){\strut{} 140}}%
\put(17191,550){\makebox(0,0){\strut{} 160}}%
\put(14950,9675){\makebox(0,0)[r]{\strut{}$b_1$ (1)}}%
\put(14950,9125){\makebox(0,0)[r]{\strut{}$b_2$ (2)}}%
\put(14950,8575){\makebox(0,0)[r]{\strut{}$\phi_1+\phi_2-4$ (3)}}%
\put(14950,8025){\makebox(0,0)[r]{\strut{}$\phi_2-3$ (4)}}%
\end{picture}%
\endgroup
\caption{Variation of $b_1$, $b_2$ and phases as function of $v$.\label{abs2}}
\end{figure}
%GNUPLOT: LaTeX picture with Postscript
\begin{figure}
\begin{picture}(0,0)%
\includegraphics{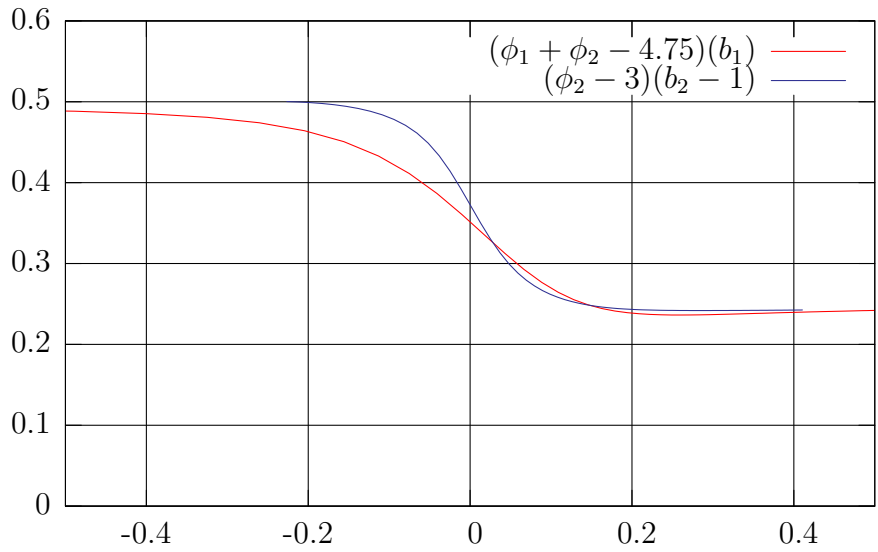}%
\end{picture}%
\begingroup
\setlength{\unitlength}{0.0200bp}%
\begin{picture}(18000,10800)(0,0)%
\put(1650,1100){\makebox(0,0)[r]{\strut{} 0}}%
\put(1650,2625){\makebox(0,0)[r]{\strut{} 0.1}}%
\put(1650,4150){\makebox(0,0)[r]{\strut{} 0.2}}%
\put(1650,5675){\makebox(0,0)[r]{\strut{} 0.3}}%
\put(1650,7200){\makebox(0,0)[r]{\strut{} 0.4}}%
\put(1650,8725){\makebox(0,0)[r]{\strut{} 0.5}}%
\put(1650,10250){\makebox(0,0)[r]{\strut{} 0.6}}%
\put(3450,550){\makebox(0,0){\strut{}-0.4}}%
\put(6500,550){\makebox(0,0){\strut{}-0.2}}%
\put(9550,550){\makebox(0,0){\strut{} 0}}%
\put(12600,550){\makebox(0,0){\strut{} 0.2}}%
\put(15650,550){\makebox(0,0){\strut{} 0.4}}%
\put(14950,9675){\makebox(0,0)[r]{\strut{}$(\phi_1+\phi_2-4.75)(b_1)$}}%
\put(14950,9125){\makebox(0,0)[r]{\strut{}$(\phi_2-3)(b_2-1)$}}%
\end{picture}%
\endgroup
\caption{Phases as a function of the roots $b_i$. \label{rel2}}
\end{figure}
We further study a case with different monodromies and values 
$n=1$ and $m=3$ for the number of zeros in each of the subintervals.
Precisely in our example we choose monodromy -1 at 0 and 1 and monodromy +1
at y=2, and obtain the plot (Figure \ref{abs2}).
In this case, the additional phases decrease from $\pi/2$ to $\pi/4$ through
$3\pi/8$ in the transition region. This difference of behavior with the case of
monodromies +1 perfectly explains the lifting of degeneracy which occurs at
small $v$ between the two degenerate levels at large $v$.
The relative plot (Figure \ref{rel2}) shows that in this case
the transition is smoother, which will be explained in the next section.

\section {Theoretical study of the transition region.}

\begin{figure}
\hfil\includegraphics[width=8cm,height=8cm]{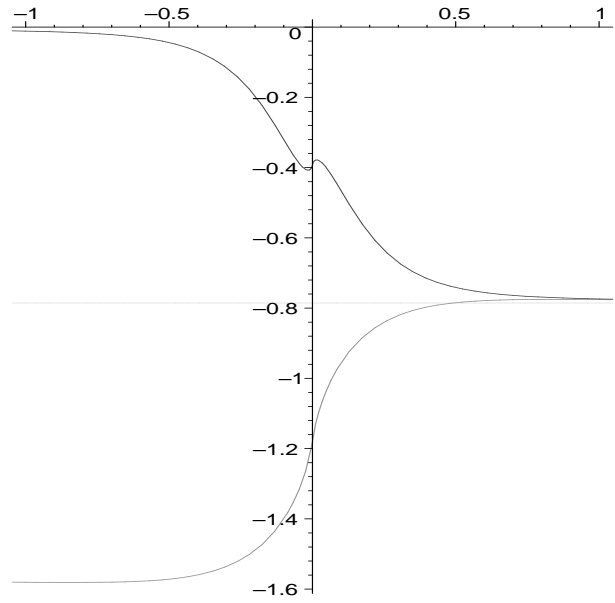}
\caption{Variation of the phases with respect to $F$.\label{phases}}
\end{figure}

To understand the behavior of the phases in the transition region we
consider for example the region around the singularity t=0 and
formulate a simplified version of eq.(\ref{wangerin}) for the case of monodromy
1 or of eq.(\ref{wangerin2}) for the case of monodromy -1. The simplified model
uses that the terms having poles at t=1 or t=y will not vary much as long as 
t remains close to 0, and doesn't approach the next pole 1. Hence we simply
replace these polar terms by constants, and study the equations, where
$f_1=v-f_2-f_3$:
\begin{eqnarray}
&&\biggl( {d^2\over dt^2} +\Bigl({1\over 2 t} + A
\Bigr)  {d\over dt} -\Bigl({f_1\over t} + B
\Bigr) \biggr) \Psi(t) = 0 \label{simpwang}\\
&&\biggl( {d^2\over dt^2} +\Bigl({3\over 2 t} + A
\Bigr)  {d\over dt} -\Bigl({f_1 - A /2
\over t} + B
\Bigr) \biggr) \Psi(t) = 0 \label{simpwang2}
\end{eqnarray}
Equations of this form can be readily related to the confluent hypergeometric
differential equation, since there is clearly a regular singularity at $t=0$
and an irregular one at $t=\infty$.

Recall the differential equation satisfied by the confluent hypergeometric
functions~\cite{AbSte}:
\begin{equation}
z {d^2w \over dz^2}+(b-z){dw \over dz} -aw=0
\label{chyperg}
\end{equation}
The solution analytic at the regular singularity $z=0$ is:
$$ M(a,b,z)=1+{az \over b} + {a(a-1) z^2 \over 2! b(b-1)} + \cdots$$
One can find its asymptotics at the irregular singularity $z=\infty$
by relating it to the Whittaker functions which have simple asymptotics there,
through the Mellin-Barnes transformation~\cite{WW}. The result is given~\cite
{AbSte} by;
\begin{equation}
{M(a,b,z) \over \Gamma(b)} = 
{e^{\pm i\pi a} z^{-a} \over \Gamma(b-a)}\Big(1+O({1\over z})\Big) +
{e^{z} z^{a-b} \over \Gamma(a)}\Big(1+O({1\over z})\Big)
\label{Masymp}
\end{equation}
where the plus sign is taken in the sector $-\pi/2<\arg z<3\pi/2$ and the minus
sign in the sector $-3\pi/2<\arg z\leq -\pi/2$.

To get eqs.(\ref{simpwang},\ref{simpwang2}) into the form eq.(\ref{chyperg})
we take $\Psi(t)=\exp(\alpha t) w(z)$ and $t=\beta z$ and observe one gets
exactly the confluent hypergeometric equation taking:
$$\alpha= -{A\over 2}+i\sqrt{B-{A^2\over 4}} ,\quad \beta= 
{i\over\sqrt{4B-A^2}}$$
Then one obtains, starting from eq.~(\ref{simpwang}) that:
$$\quad a= {1 \over 4} -\beta(f_1 - {1\over 4}A)  ,\quad b= {1\over 2}   $$
and similarly starting from eq.~(\ref{simpwang2}) that:
$$ a= {3\over 4}-\beta(f_1- {5 \over 4}A) 
,\quad b= {3\over 2}$$
Finally remembering that we need precisely the solution which is regular at t=0
in both cases, the corresponding solution is thus $M(a,b,z)$, whose asymptotic
expansion we know from eq.(\ref{Masymp}), and which can be directly compared
with the semi--classical approximation. Of course the validity of this
procedure rests on the hypothesis that $t$ is large enough so that one can
apply the asymptotic analysis, still small enough that the next pole,
$t=1$ remains distant. This is the case when $\beta$ is small, and is
thus justified in our case for large $B$, that is when $f_2$ and
$f_3$ are large enough. Note that in this case $A$ is small compared to
$B$.  In the two cases, $a$ takes the form
${1\over 4} -i F $ or ${3\over 4} -iF'$.

Applying the asymptotic formula~(\ref{Masymp}) we get for large positive
$t$, respectively:
\begin{eqnarray}
 \Psi(t) &\simeq& {1\over t^{1/4}} \cos \bigl( \omega t - {\pi
\over 8 } +F \ln (2\omega t) - \Im \ln \Gamma( {1\over4}+iF ) \bigr)\\
 \Psi(t) &\simeq& {1\over t^{3/4}} \cos \bigl( \omega t - {3\pi \over 8 }
+F' \ln (2\omega  t) - \Im \ln \Gamma( {3\over4}+iF' ) \bigr)
\end{eqnarray} 
with $\omega = \sqrt{B-A^2/4}$. We then have to compare these results with the
phase we obtain from the WKB analysis of eqs.~(\ref{simpwang}), which covers
the cases of the two monodromies. The semiclassical action is given by:
$$
\int_0^t \sqrt{B t' + f_1 \over t' } dt' = \sqrt{t(Bt+f_1)} +{f_1 \over 2 
\sqrt B} \ln \bigl(1 + 2 B t /f_1 + 2 \sqrt{ ( 1 + Bt/f_1)(Bt/f_1) } \bigr)
$$
In the limit of large $Bt/f_1$, this is simply
$$\omega t + F \ln(2\omega t) + F - F \ln(F)$$
with the same quantities $\omega$ and $F$ defined above, but taken with $A=0$.
If $A=0$, we deduce that the phase difference between the semiclassical result
and the exact one are therefore:
\begin{eqnarray}
\Delta \phi_+ &=& {\pi\over 8} + \Im \ln \Gamma( {1\over4}+iF ) -F\ln(F) + F\\
\Delta \phi_- &=& {3 \pi\over 8} + \Im \ln \Gamma( {3\over4}+iF' ) -F'\ln(F')
+F'
\end{eqnarray} 
In fact, $F=F'$ when $A=0$. The variation of the phases $\Delta \phi_\pm$ are
plotted in (Figure \ref{phases}), which shows that this analytical study
recovers the limit of the phases found numerically in the preceding section.
This plot shows a striking similarity with the above numerical results, complete
with the `glitch' around 0 in $\Delta \phi_+$, which comes from the singularity
of the derivative of $F\ln(F)$ at $F=0$. One could worry on the possibility to make 
a proper match when $A$ is non zero. However, the difference on $\omega$ is quadratic in $A$
and is further divided by $\sqrt B$, so that the corresponding phase difference cannot become
significant over the range necessary to obtain a proper match. The term proportional to the logarithm
of $\omega t$ is linear in $A$, but the slow variation of the logarithm makes it also benign, so that the 
only contribution which is significant is in the phase of the $\Gamma$ function. This should explain 
the fact that when the $b_{i}$ cross their critical line, which corresponds to $f_{1}=0$, we obtain 
phases which are different from the $\pi/8$ or $3\pi/8$ which we  obtain in the simpler treatment 
with $A=0$. Changing the sign of $t$ corresponds to keeping $B$ while changing the signs of $f_{1}$ and $A$.  Our formulae for the $\Delta \phi_{\pm}$ show that the phases on both side of a singularity 
add up to $\pi/4$ or $3\pi/4$ as we supposed in section 2 for the invariance of the semiclassical 
energy at $v=0$

\section {Conclusion.}

The combined results of~\cite{BT05a} and the present work show that it is
possible to reproduce the entire spectrum of the quantum Neumann model with
semiclassical methods in the whole range of parameters. The accuracy remains
good even for low lying levels, but finite differences remain between the
semiclassical and exact results. The new feature in the realm of semiclassical
studies is that the known facts about the asymptotic behaviour of confluent
hypergeometric functions allows to properly model the transition region between
different types of boundary conditions. This introduces contributions which
cannot be reduced to a discrete Maslov index.


\begin{thebibliography}{10}


\bibitem{Gurarie}
D.~Gurarie, {\em Quantized Neumann problem, separable potentials on S$^n$ and 
the Lam\'e equation.} 
\newblock J. Math. Phys. 36 (1995), pp. 5355--5391.


\bibitem{BT05}
M. Bellon and M. Talon,
\newblock {Spectrum of the quantum Neumann model.}
\newblock Phys. Lett. A 337 (2005), pp. 360--368.

\bibitem{BT05a}
M. Bellon and M. Talon,
\newblock {The quantum Neumann model: asymptotic analysis.}
\newblock hep-th/0507207.

\bibitem{Pau}
W. Pauli,
\newblock {\em Pauli lectures on physics 5 Wave mechanics.}
\newblock {MIT Press, 1973.}

\bibitem{Lan38}
Rudolph E.\ Langer, {\em On the Connection Formulas and the
Solutions of the Wave Equation.} \newblock Phys.\ Rev.\ {\bf 51}(1938),
pp.\ 669--676.


\bibitem{WW}
E.T. Whittaker and G.N. Watson.
\newblock {\em A course of modern analysis}.
\newblock Cambridge University Press,  (1902).

\bibitem{AbSte}
M. Abramowitz and I. Stegun
\newblock {\em Handbook of mathematical functions.}
\newblock National Bureau of Standards 1964.



\end{thebibliography}
\end{document}